# Methods for The Testing of Nanopowders

V.V. An[1], Ch. de Izarra[2], A.V. Korshunov[1], A.Yu. Godimchuk[1], Yu.A. Amelkovich[1], G.V. Yablunovskii[1]

1- High Voltage Research Institute, Tomsk Polytechnic University, 2a Lenin Ave., Tomsk 634050, Russia

2 -LASEP Faculté des Sciences, Site de Bourges, Université d'Orléans, rue Gaston Berger, F-18028 Bourges Cedex

(France)

Nowadays nanopowders of metals, alloys, intermetallides, and their chemical compounds are produced by different ways. One of the problems suppressing promotion of nanopowders to the industry is the absence of reliable and objective methods for their testing. Undoubtedly, an important moment in testing is a correct interpretation of obtained experimental data.

The objective of this work is an overview of existing testing methods, analysis of these methods adequacy, and estimation of a possibility to create a system for nanopowders testing.

In the work, nanopowders produced by wire electric explosion (WEE) were used for experiments [1]. Nanopowders produced by evaporation-condensation in gas media and industrial powders with a diameter of 1-2 μm were also used for comparison.

The most important parameters, which give a possibility to refer an analyzed powder to nanopowders, are characteristic sizes of particles ($\bar{a}_0$) and specific surface area ($S_{sp}$). In case of irregular particles shape, it is necessary to take into account 3 characteristic values: length, width, and thickness. The integral characteristic is $S_{sp}$ and conditional surface-average diameter of particles ($\bar{a}_s$):

$$\bar{a}_s = \frac{6}{\rho \cdot S_{sp}} \qquad (1)$$

Measurement accuracy of $S_{sp}$ depends strongly on the porosity of particles (surface smoothness). Open porosity leads to the understated particles diameter, whereas closed porosity gives the overstated one. Thus, the limit between nanoparticles and sublimated particles is very largely uncertain.

When analyzing the main component content in nanopowders, difficulties in obtaining reliable results arise. It is connected with the presence of gases, water, oxide-hydroxide compounds, carbonates, and absorbed gases, e.g., hydrogen. According to the experimental estimation, the total content of gaseous impurities in nanopowders may be up to 7-8 wt%. Experimental results show that it is necessary to vacuumize the nanopowders (not more than 10 Pa) and to treat it with helium at 70-90°C in order to remove adsorbed gases and loosely coupled chemical compounds. Absorbed gases are removed at higher temperatures (400-600°C) that, in most cases, lead to nanopowders sintering. The synthesis and preparation of composition materials with nanopowders can lead to "apparent violation" of material balance because of removing up to 79 wt% of gaseous and chemically loosely coupled substances handling by nanosystems (solid – gas).

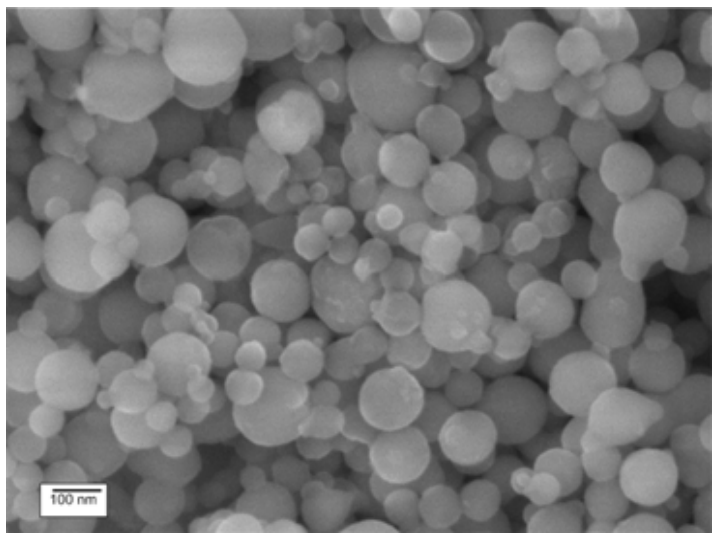

Fig.1. SEM image of Al nanopowders produced by WEE.

Particularities of an analyzed object should be taken into consideration for X-ray analysis of nanopowders. At low angles of X-ray irradiation on a sample, a large halo is observed for angles 2θ of 6-10°. This is identified with amorphous components of substances, areas of coherent scattering ($D_{ACS}$) etc. As is well known for powder samples, an optimum – maximal reflection of X-ray irradiation is observed in case of characteristic sizes of particles in the range of 100-500 nm.

Normally, studied nanopowders are polydisperse (Fig.1). The fraction of particles < 100 nm increases the background response, promotes a halo, leads to reflections enlargement. The fraction of particles >50 μm gives symmetry and distorts reflections (Fig.2). X-ray analysis is particularly complicate for composition lamellar particles.

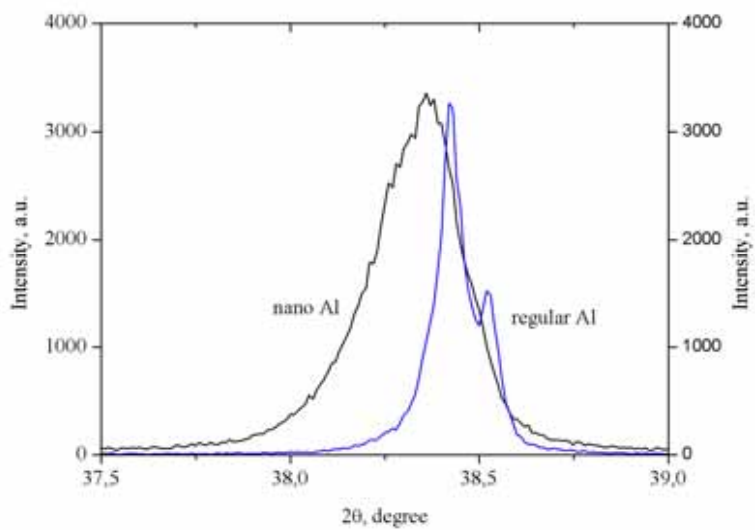

Fig.2. X-ray patterns of nanosized and regular Al.

For heavy metals, the information about the composition of nanoparticles can be distorted. For instance, the analysis of tungsten carbide nanopowders has shown that it consists of the $WC_{1-x}$ phase. According to analysis of carbon content, the essential part in the particle is non-oxidized α-W. When etching the surface with argon plasma, the α-W phase was detected in the center of particles after removing tungsten carbide. For heavy metals, the depth of X-ray penetration is several tens of angstroms. In case of composition particles analysis, the shielding of one substance by another should therefore be taken into account even for semiqualitative estimation of X-ray data.

**Parameters of nanopowders reactivity.** The transfer of substances into the nanostate requires power inputs which partially remain in the substance. The nanostate is therefore a metastable or unstable state. It is expedient to test reactivity of nanopowders. In this case, the differential thermal analysis and the differential scanning calorimetry with the record of thermogravimetry are handy and informative methods. The feature of nanopowders analysis is always connected with a probability of their chemical and diffusion interaction with diagnostic cell materials. Nanopowders of aluminum, iron, nickel and other metals alloy with platinum, corundum, alundum and tantalum during heating in air. The nanopowders reactivity is supposed to be estimated quantitatively using 4 parameters (Tab.1, Fig.3):

- temperature of the onset of oxidation ($T_{on}$, °C);
- maximum oxidation rate ($v_{ox}$);
- degree of aluminum conversion (α, %);
- ratio of the thermal effect to the weight increase measured under standard conditions (S/Δm) [2].

Table 1. Parameters of Al powders reactivity

| Sample | $S_{sp.}$(BET), $m^2/g$ | $ā_s$, μm | $C_{Al}^0$, % | $T_{on}$, °C | α, % | | $v_{ox}$, mg/s (in a temperature range, °C) | S/Δm, rel.un. |
|---|---|---|---|---|---|---|---|---|
| | | | | | up to 660°C | up to 1000°C | | |
| 1 | 0.38 | 9.0 | 98.5 | 820 | 2.5 | 41.8 | 0.05 (970-980) | - |
| 2 | 16.00 | 0.13 | 89.0 | 540 | 50.1 | 78.6 | 0.05 (550-605) | 8.7 |

1 – Regular Al;
2 – Nanosized Al produced by WEE.

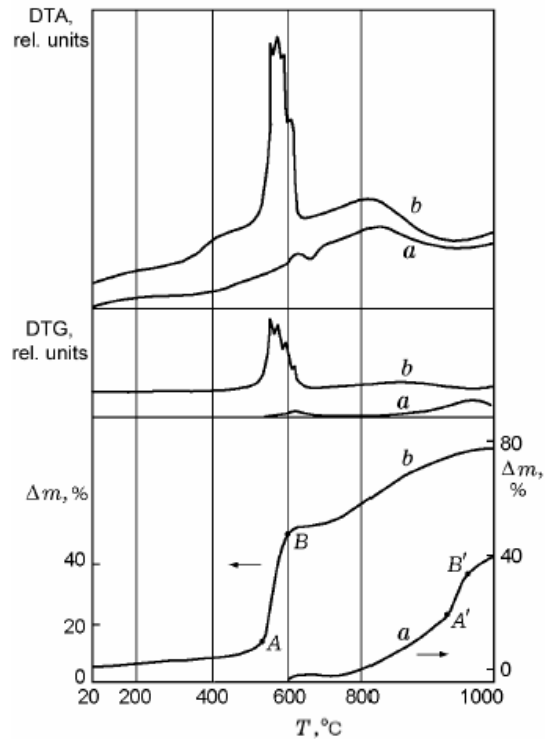

Fig.3. DTA, DTG, and TG of regular (a) and nanosized aluminum (b).

**Enthalpy of nanopowders formation.** For an integral estimation of nanopowders state, a diagnostic parameter $\Delta H_f^0$, which is indirectly connected with reactivity and diffusion activity of nanopowders, is necessary. When dispersing solids, an increase of structural units (SU) occurs on the surface. Their part can be comparable with a part of SUs (atoms) in the volume of particles. The SUs on the surface reveal oscillation amplitudes higher than those in the volume. Consequently, they possess a higher stored energy. Point defects can contribute to the stored energy of a single particle, whereas a set of particles (powder) has also a surface energy. The stored energy can release during combustion, dilution, sintering and other processes. Thus, the thermodynamic state of nanopowders is characterized by a positive enthalpy of formation ($+\Delta H_f^0$) and increased entropy. Without taking into account the enthalpy factor, the enthalpy of nanopowders was determined using qualitative differential thermal analysis (DTA).

A derivatograph Q-1500 was used for measurements of the heat released during heating of nanopowders in air. Differential thermal analysis (DTA) is quite informative. However, processes of heat transfer, heat irradiation and the use of reference substances should be taken into consideration in case of qualitative measurements.

Results of ΔH measurements for 4 samples of aluminum nanopowders are given in Tab.2.

Table 2. Results of differential thermal analysis of Al nanopowders.

| Parameter | Al nanopowder sample | | | |
|---|---|---|---|---|
| | 1 | 2 | 3 | 4 |
| Mass of a sample $m_s$, $\times 10^{-3}$ g | 47,2 | 59,9 | 54,4 | 57,4 |
| Temperature range of heat release, °C | 507÷643 | 496÷699 | 545÷699 | 485÷692 |
| $\Delta H^{sp}_{exp.en.}$ kJ/mole | +5±5 | +10±5 | +11±7 | +5±7 |

The obtained results are characterized by a significant value dispersion that allows only a qualitative estimation of the excess energy stored in nanopowders.

**Conclusion**

For nanopowders testing, the following characteristics are proposed to be measured: main substance content, wt%, characteristic sizes of particles $ā_s$, specific surface area, phase composition (X-ray phase analysis), substructural parameters (X-ray structure analysis), enthalpy of formation $\Delta H_f^0$ using DTA and parameters of metal nanopowders reactivity at their heating in air.

**References**

1. Y.-S. Kwon, Y.-H. Jung, N.A. Yavorovsky, A.P. Illyn and J.-S. Kim. Ultra-fine powder by wire explosion method. Scripta mater. 44 (2001), pp.2247–2251.
2. A. Ilyin, A. Gromov, V. An, F. Faubert, Ch. de Izarra et al. Characterization of Aluminum Powders I. Parameters of Reactivity of Aluminum Powders. Propellants, Explosives, Pyrotechnics 27 (2002), pp.361-364.